\documentclass[aps,prd,twocolumn,10pt,superscriptaddress,footinbib,nobibnotes,nolongbibliography]{revtex4-2}

\usepackage{amssymb}
\usepackage{graphicx}
\usepackage{amsmath}
\usepackage{hyperref}
\usepackage{multirow}
\usepackage{booktabs}
\usepackage{setspace}
\usepackage{verbatim}
\usepackage{float}
\usepackage[normalem]{ulem}
\usepackage{arydshln}
\usepackage[table,xcdraw]{xcolor}
\usepackage{subcaption}
\usepackage{lipsum}
\usepackage{physics}
\usepackage{etoolbox}
\usepackage{comment}
\usepackage{soul}

\setstcolor{red}  

\usepackage[english]{babel}
\usepackage[utf8]{inputenc}
\usepackage[colorinlistoftodos, color=green!40, prependcaption]{todonotes}

\def\Msun{{\rm M}_{\odot}}

\begin{document}
\title{Dwarf Galaxy Reticulum II's MeerKAT L-Band Observation: Constraining WIMP Dark Matter}

\author{Phillip Badenhorst}
    \affiliation{Department of Physics, Stellenbosch University, Matieland 7602, South Africa}

\author{Michael Sarkis}
    \affiliation{Department of Physics, Stellenbosch University, Matieland 7602, South Africa}

\author{Yin-Zhe Ma}
\email{Corresponding author: mayinzhe@sun.ac.za; South Africa Research Chair (SARChI)}
    \affiliation{Department of Physics, Stellenbosch University, Matieland 7602, South Africa}

\author{Wen-Qing Guo}
    \affiliation{Department of Physics, Stellenbosch University, Matieland 7602, South Africa}

\author{Geoff Beck}
    \affiliation{School of Physics and Centre for Astrophysics, University of the Witwatersrand, Johannesburg, South Africa, WITS-2050}

\author{Qiang Yuan}
    \affiliation{Key Laboratory of Dark Matter and Space Astronomy, Purple Mountain Observatory,
Chinese Academy of Sciences, Nanjing 210023, China}

\author{Gyula I.G. J\'{o}zsa}
    \affiliation{Max-Planck-Institut f\"{u}r Radioastronomie, Auf dem H\"{u}gel 69, 53121 Bonn, Germany}
    \affiliation{Centre for Radio Astronomy Techniques and Technologies (RATT), Department of Physics and Electronics, Rhodes University, Makhanda 6140, South Africa}

\author{Steven G. Murray}
    \affiliation{Department of Physics, Stellenbosch University, Matieland 7602, South Africa}
    

\begin{abstract}
The nature of dark matter remains one of the most profound open questions in modern physics, with weakly interacting massive particles (WIMPs) among the leading candidates proposed to account for the observed anomalies across cosmic scales. If WIMPs annihilate or decay in the dense dark matter haloes of nearby galaxies, the resulting high-energy electrons/positrons can produce synchrotron radiation in the presence of ambient magnetic fields, yielding a potentially detectable radio signature. Dwarf spheroidal galaxies are particularly promising targets for indirect dark matter searches due to their high mass-to-light ratios and low astrophysical foreground contamination. While radio observations have been widely used to constrain WIMP properties in these systems, the resulting limits remain highly sensitive to poorly constrained magnetic field assumptions. Here we utilize L-band ($1.3$--$1.5$\,GHz) observations from the MeerKAT array directed towards dwarf galaxy Reticulum II and measure the surface brightness profile across the dwarf galaxy. Combining these measurements with theoretical models of WIMP annihilation and yield 95\% confidence level upper limits on the WIMP annihilation cross-section $\langle\sigma v\rangle \lesssim 10^{-26}$ cm$^3$ s$^{-1}$ for masses $m_\chi \sim 10$--$200$\,GeV (with the assumed uniform magnetic field model and diffusion constant). As next-generation radio facilities such as the Square Kilometre Array come online, robust characterisation of magnetic environments in dwarf galaxies will be essential to fully exploit their sensitivity to the particle nature of dark matter.
\end{abstract}

\keywords{dark matter, dwarf spheroidal galaxies, astroparticle physics, radio continuum: galaxies, magnetic fields, galaxies: individual: Reticulum II}

\maketitle

\section{Introduction} \label{sec:intro}
Unraveling the underlying composition of dark matter (DM) stands as a central pillar of contemporary astrophysics and particle physics. Although its existence is firmly established through a wide range of gravitational observations, including temperature anisotropies in the cosmic microwave background (CMB) \cite{Planck:2018vyg}, galaxy rotation curves \cite{1970ApJ...159..379R,ES2026}, and gravitational lensing \cite{Massey_2010,Qu2026}, the particle nature of DM remains unknown. Among the various particle candidates proposed, weakly interacting massive particles (WIMPs) are particularly well motivated, as they naturally arise in extensions of the Standard Model and can account for the observed DM abundance through thermal freeze-out \cite{Roszkowski:2017nbc, Arcadi:2017kky}.

WIMPs can be searched for through three complementary approaches \cite{Arcadi:2017kky, cirelli2025darkmatter}: collider experiments \cite{Giagu:2019fmp}, direct-detection \cite{Schumann:2019eaa}, and indirect detection \cite{Hooper:2010mq, Huang:2011xr, Funk:2013gxa, Conrad:2014tla, Fermi-LAT:2015att, Klasen:2015uma, Gaskins:2016cha, Leane:2020liq, DiMauro:2023qat}. Collider searches aim to produce DM particles in high-energy collisions, while direct-detection experiments seek to observe their scattering with nuclei or electrons, such as the recent LUX-ZEPLIN results~\cite{Akerib2026}. Indirect detections instead target the Standard Model (SM) particles produced by DM annihilation or decay, such as gamma rays and neutrinos. These complementary approaches probe different aspects of the underlying DM interactions and therefore provide independent constraints on WIMP properties~\cite{Guo:2022rqq,Semane_2026}.

In addition to gamma rays and neutrinos, WIMP annihilation or decay can produce high-energy electrons and positrons. In the presence of magnetic fields, these relativistic particles lose energy through synchrotron radiation, producing emission primarily at radio frequencies. Radio observations therefore provide a complementary probe of WIMP-induced diffuse radio emission \cite{Colafrancesco:2006he,Spekkens:2013ik,Natarajan:2013dsa, Colafrancesco:2014coa,Beck:2015rna,Regis_2017,Kar:2019hnj,Vollmann:2019boa, Bhattacharjee:2019jce, Bhattacharjee:2020phk, Kar:2020coz, Cembranos_2020, Guo:2022rqq, Chen:2021rea, Regis:2021glv,Basu:2021zfg, Semane_2026,regis2026probingfundamentalnatureparticle}. With the increasing sensitivity of modern radio interferometers, such as the MeerKAT array \cite{Jonas2016MeerKAT}, radio observations have become a promising avenue for indirect DM searches.

In terms of the targets of search, nearby dwarf spheroidal (dSph) galaxies are particularly attractive targets. Their proximity allows relatively sensitive observations of faint extended emission, while their large mass-to-light ratios indicate that they are strongly DM dominated. At the same time, dSph galaxies generally lack substantial conventional astrophysical emission, providing relatively clean environments in which to search for a DM-induced signal \cite{2004PhRvD..69l3501E, 1990Natur.346...39L}. Furthermore, the presence of interstellar magnetic fields at the $\mu$G level can facilitate synchrotron emission from relativistic electrons and positrons produced by DM annihilation or decay. These properties make nearby dSph galaxies promising laboratories for complementary radio searches for particle DM.

Of particular interest is the nearby dSph galaxy Reticulum II (RetII), located at a distance of approximately $31.2\,{\rm kpc}$ and with a measured mass-to-light ratio of $470\pm210$ $(\text{M}_{\odot}/\text{L}_{\odot})$ \cite{Simon_2015}. Tentative gamma-ray excesses have been reported from the direction of RetII \cite{Geringer-Sameth:2015lua}, although their origin remains uncertain. If such an excess were associated with WIMP annihilation or decay, the same process would generically produce energetic electrons and positrons, and hence an associated synchrotron signal in the presence of a magnetic field. A radio search therefore provides an independent and complementary test of a possible DM interpretation of the gamma-ray emission. In this work, we observed RetII for a total time of 6 hours using the L-band of MeerKAT to investigate the potential DM-induced synchrotron signals. We consider both WIMP annihilation and decay scenarios and investigate how the predicted radio emission depends on the assumed magnetic-field model as well as the diffusion properties of the relativistic electrons and positrons. 

The remainder of this paper is organized as follows. Section~\ref{sec:reduction} describes the data reduction and calibration of the MeerKAT observations. Section~\ref{sec:model} presents the DM signal model, including the magnetic-field configurations and particle diffusion. Section~\ref{sec:likelihood} details the statistical likelihood and spatial covariance methodology. Section~\ref{sec:results_discussion} presents the resulting constraints on WIMP annihilation and decay scenarios. Finally, Section~\ref{sec:conclusions} summarizes our main conclusions.

\section{Observational Data} \label{sec:reduction}
\subsection{MeerKAT and Observation}

MeerKAT is a 64-antenna radio dish array located in the Karoo desert of South Africa. Each dish is 13.5 meters in diameter, with baseline lengths from 29--7,700\,m \cite{sarao_meerkat_specs}. The array's geographic center is located at $30^\circ 42' 39.8''$ South, $21^\circ 26' 38.0''$ East, at an elevation of 1,086.6\,meters above sea-level. 
The sensitivity of the observations depends on the system temperature, consisting of all factors that influence the signal path:
\begin{equation}
    \begin{split}
        T_{\text{sys}}(\vec{\theta},\nu) =& \hspace{0.3em} T_{\text{sky}}(\vec{\theta}, \nu)+T_{\text{spill}}({\it el},\nu) \\
        &+T_{\text{atmo}}({\it el}, \nu)+T_{\text{rec}}, \label{eq:T_sys}
    \end{split}
\end{equation}
parameterized by dish elevation {\it el}, pointing direction $\vec{\theta}$, and observational frequency $\nu$. The sky temperature is inherently a direction-dependent quantity. For observations away from the Galactic plane, the effective sky temperature can be accurately approximated by a ``cold'' patch of sky as $T_{\text{sky}}[{\rm K}] \approx 2.725 + 1.6(\nu/\text{GHz})^{-2.75}$ \cite{Santos:2015gra}, comprising the CMB baseline and high-latitude Galactic synchrotron emission. RetII is located at high Southern Galactic latitude, thus the diffuse Galactic emission is minimal, and this cold-sky approximation holds exceptionally well (yielding $T_{\text{sky}} \approx 3.4$\,K across MeerKAT's L-band, where the total $T_{\text{sys}} \approx 18\text{--}20$\,K is predominantly set by receiver noise \cite{Mauch_2020}). The atmospheric and ground spillover contributions depend on elevation angle and atmospheric opacity, while the receiver temperature characterizes the thermal noise from the feed horn, the cooled low-noise amplifiers, and front-end analog electronics.
The sensitivity of an individual antenna is characterized by its System Equivalent Flux Density (SEFD), defined as
\begin{equation}
    \text{SEFD}(\nu) = \frac{2k_{\rm B} \, T_{\text{sys}}(\nu)}{A_{\text{eff}}(\nu)},
    \label{eq:sefd}
\end{equation}
where $k_{\rm B}$ is the Boltzmann constant, and $A_{\text{eff}}(\nu) = \eta_{\text{ap}}(\nu) A_{\text{geom}}$ is the effective collecting area of the dish, scaled from its geometric area ($A_{\text{geom}} \approx 143.1\text{ m}^2$) by the frequency-dependent aperture efficiency $\eta_{\text{ap}}(\nu) \approx 0.70\text{--}0.78$ across L-band \cite{sarao_meerkat_specs, Jonas2016MeerKAT}.

For an interferometric array of $N_{\text{ant}}$ identical antennas with two orthogonal polarizations, the theoretical $1\sigma$ thermal noise floor in a naturally-weighted Stokes $I$ synthesized image is governed by the interferometric radiometer equation \cite{Thompson2017}:
\begin{equation}
    \sigma_{\text{RMS}} = \frac{\text{SEFD}}{\eta_{\text{c}} \sqrt{2 \Delta\nu \, t_{\text{int}} \, N_{\text{ant}}(N_{\text{ant}}-1)}},
    \label{eq:radiometer_sensitivity}
\end{equation}
where $\Delta\nu$ is the synthesized bandwidth (either per spectral channel $\Delta\nu_{\text{ch}}$ or across the full continuum bandwidth), $t_{\text{int}}$ is the total on-source integration time, $\eta_{\text{c}} \approx 0.85\text{--}0.95$ represents the digital correlator efficiency, and $N_{\text{ant}}(N_{\text{ant}}-1) = 2 N_{\text{bl}}$ accounts for the $N_{\text{bl}} = 2016$ cross-correlated baselines formed by the 64 dishes.

\begin{figure*}[htbp]
        \centering
        \includegraphics[width=\textwidth]{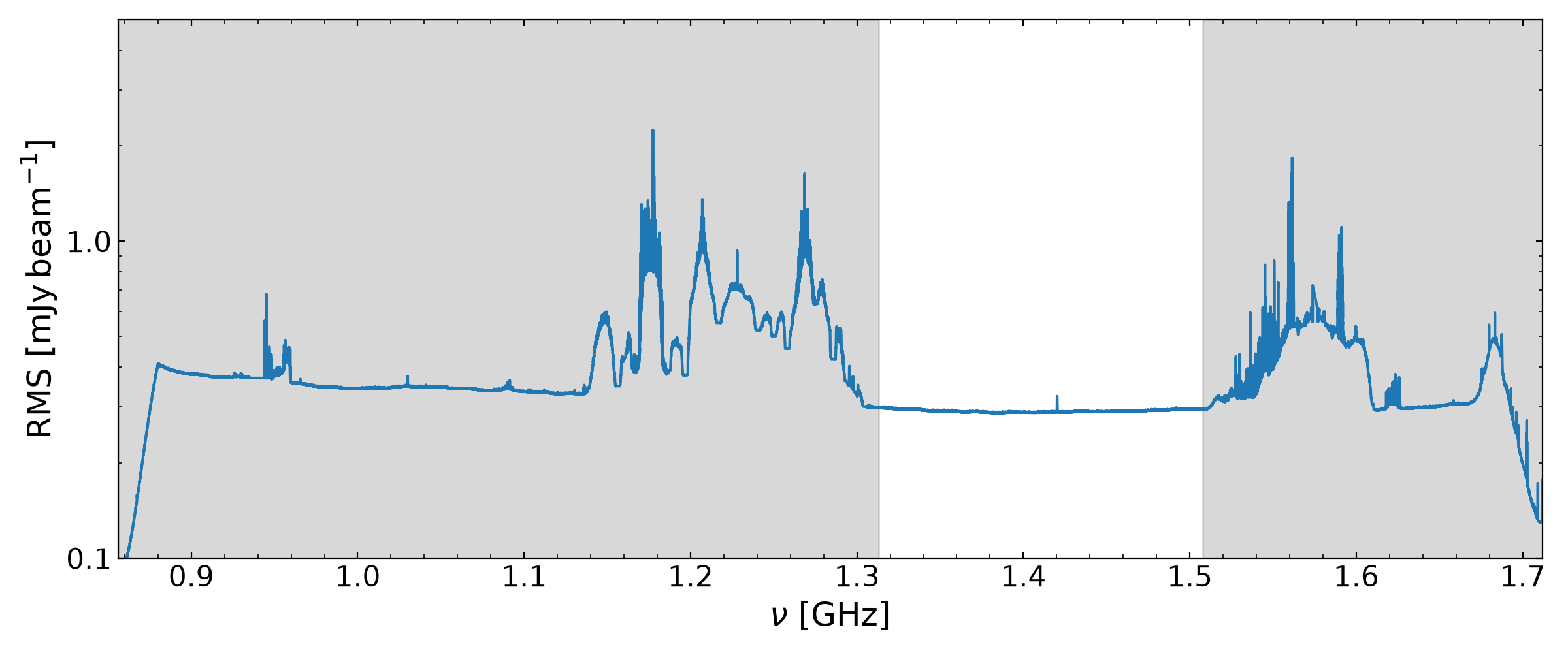} 
        \caption{The per-channel RMS noise profile ($\sigma_{\text{RMS}}$) measured across the full MeerKAT L-band after cross-calibration. In RFI-free channels, the empirical noise approaches the theoretical radiometer limit ($\approx 0.3\text{ mJy beam}^{-1}$). The shaded grey regions designate frequency ranges contaminated by persistent satellite and terrestrial transmissions, which are ultimately excised from the analysis. The unshaded white region (1313--1508\,MHz) marks the cleanest observing band retained for our science analysis. Automated visibility-level RFI flagging within this band was performed using \texttt{Tricolour}, and the narrow spike at $\approx 1420$\,MHz corresponding to Galactic HI line emission was explicitly masked.}
        \label{fig:full_rms}
    \end{figure*}

To evaluate the actual noise performance across the observation, Fig.~\ref{fig:full_rms} presents the per-channel noise profile, $\sigma_{\text{RMS}}(\nu)$, measured across the full MeerKAT L-band (856--1712\,MHz) following cross-calibration. In the absence of radio frequency interference (RFI), the measured noise floor approaches the theoretical thermal limit predicted by Eq. \ref{eq:radiometer_sensitivity} ($\sim 0.3\text{ mJy beam}^{-1}$ per channel). However, prominent spikes in the measured RMS are evident across several frequency intervals due to persistent terrestrial and satellite transmissions below $\sim 1.3\text{ GHz}$, as well as GPS mobile satellites above $\sim 1.5\text{ GHz}$. 

RetII has been an observational target of choice for a multitude of DM searches across various DM models \cite{Bonnivard_2015, PhysRevD.102.062001, Duerr_2016}. However, the very characteristics that make ultra-faint dwarf galaxies ideal targets also make determining their astrophysical parameters inherently challenging. Currently, the most precise constraints for these parameters are provided by the Dark Energy Survey (DES) \cite{Bechtol_2015}. We provide a comprehensive summary of the properties of RetII useful to the analysis, listed in Table~\ref{tab:RetII}. In 2022, we geared the MeerKAT telescope towards RetII for a total period of 6 hours (MKT-22149), as well as a primary calibrator source J1939-6342 (RA 19:39:25, Dec -63.42.45), a secondary calibrator J0408-6545 (RA 04:08:20, Dec -65.45.09) and a phase/amplitude calibrator J0303-6211 (RA 03:03:50, Dec -62.11.25). The primary and secondary calibrators were used for bandpass and flux calibration.

\begin{table}[htbp]
    \centering
    \caption{Properties of Reticulum II.}
    \label{tab:RetII}
    \begin{tabular}{c l r}
        \hline
        \textbf{Row} & \textbf{Quantity} & \textbf{Value} \\
        (1) & RA (J2000) & 03:35:41 \\
        (2) & Dec (J2000) & -54:03:00 \\
        (3) & Distance (kpc) & $31.2$ \\
        (4) & $r_{1/2}$ (pc) & $55 \pm 5$ \\
        (5) & $M_{1/2}$ ($\text{M}_{\odot}$) & 5.6 $\pm$ 2.4 $\times 10^{5}$ \\
        (6) & $M$/$L_{\text{V}}$ ($\text{M}_{\odot}/\text{L}_{\odot}$) & 470 $\pm$ 210 \\
        \hline
    \end{tabular}
    \caption*{Rows (1)-(3) are derived from the DES photometric analysis \cite{Bechtol_2015}, while rows (4)-(6) are from the stellar kinematics and metallicity study \cite{Simon_2015}.}
    \captionsetup{font=small}
\end{table}
    
\subsection{Data Reduction}

The data reduction was done using different workers from the \texttt{CARACal} pipeline \cite{2020ascl.soft06014J}, as well as source subtraction using the Python Blob Detector
and Source Finder (PyBDSF) \cite{2015ascl.soft02007M}. Following data ingestion and initial weight verification, the multi-source observation (i.e., the primary target and calibrator targets) was partitioned into separate Measurement Sets (\texttt{ms}) for the science target and individual calibrator fields using the \texttt{mstransform} utility provided by the Common Astronomy Software Applications (CASA) \cite{The_CASA_Team_2022}. 

Prior to calibration, a preliminary static flagging pass was applied to mask instrumental bandpass edge roll-offs (discarding channels below $\approx 880\text{ MHz}$ and above $\approx 1680\text{ MHz}$ where receiver response degrades) and to excise the Galactic neutral hydrogen HI line emission window between $1419.8$ and $1421.3\text{ MHz}$. Broad automated RFI excision was subsequently executed across time and frequency domains using the \texttt{Tricolour} package, which implements the flagger based on the SumThreshold combinatorial thresholding algorithm.

Following RFI flagging, we performed first-generation (1GC) direction-independent cross-calibration using CASA \cite{The_CASA_Team_2022} within the \texttt{crosscal} worker of \texttt{CARACal}. Cross-calibration solves for instrumental and atmospheric corruptions by observing bright external reference sources alongside the science target. Antenna delays and frequency-dependent bandpass responses were solved on the primary calibrator, while time-dependent complex gain solutions were solved on the secondary calibrator observed at high cadence throughout the 6-hour track using \texttt{gaincal}. To set the absolute flux scale, CASA's \texttt{fluxscale} task scaled the gain amplitudes between both calibrators, establishing the secondary calibrator's absolute flux density and spectral index. Finally, the bandpass and flux-scaled gain solutions, which were interpolated in time across target scans, were applied to the RetII visibilities using \texttt{applycal}.

We then performed second-generation calibration (2GC), or direction-independent self-calibration, on the target dataset using the \texttt{selfcal} worker of \texttt{CARACal}. Prior to self-calibration, a secondary pass of automated flagging was executed on the target visibilities using \texttt{Tricolour} to excise low-level residual RFI. To initialize self-calibration without deconvolution bias, a clean mask was constructed from cataloged bright radio sources in the Sydney University Molonglo Sky Survey (SUMSS) \cite{10.1046/j.1365-8711.2003.06605.x}. This mask constrained the initial deconvolution to verified astrophysical sources, preventing sidelobes or noise spikes from being erroneously incorporated into the calibration sky model. The pipeline then iteratively solved for time-dependent antenna phase and amplitude gains by minimizing the residuals between the observed visibilities and the Fourier-transformed sky model. Imaging was carried out using Briggs weighting with a robustness parameter of $0.5$, which provides a balance between maximizing point-source sensitivity and suppressing Point Spread Function (PSF) noise. This self-calibration stage produced Multi-Frequency Synthesis (MFS) images of the target field, including the dirty image, the deconvolved sky model, the residual map, and the final restored image.

\begin{figure*}[htbp]
        \centering
        \includegraphics[width=\textwidth]{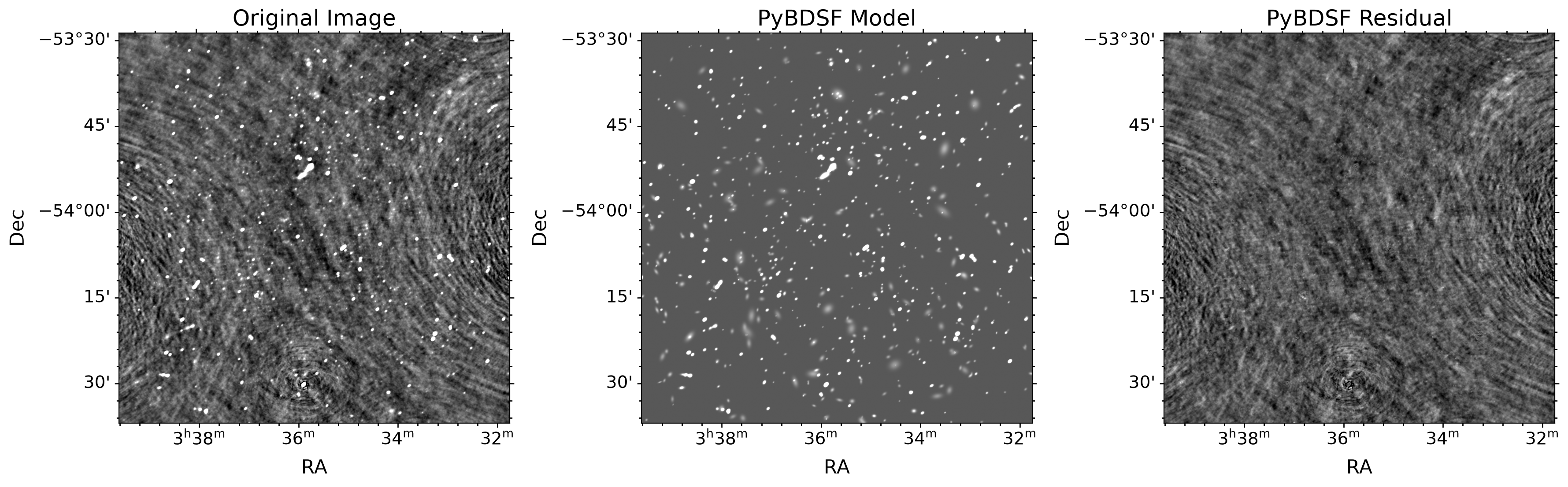} 
        \caption{PyBDSF discrete source extraction on the 2GC self-calibrated MFS image of the RetII field. \textit{Left}--The input restored image prior to source subtraction. \textit{Middle}--The PyBDSF model image reconstructed from the fitted two-dimensional Gaussian components of all detected discrete sources. \textit{Right}--The residual image resulting from directly subtracting the Gaussian source model from the restored image. Residual artifacts radiating from bright off-axis sources remain prominent, demonstrating the presence of direction-dependent phase and amplitude errors that cannot be corrected by direction-independent self-calibration alone.}
        \label{fig:pybdsf}
    \end{figure*}

Next, PyBDSF \cite{2015ascl.soft02007M} was employed to identify and model discrete background radio sources across the target field. PyBDSF computes a spatially varying background RMS noise map across a sliding two-dimensional grid, which accounts for elevated local noise near bright sources. Contiguous emission zones exceeding a set detection threshold are then decomposed into a collection of two-dimensional Gaussian components. Subtracting this model from the restored image yields a source-subtracted residual map (Fig.~\ref{fig:pybdsf}, right panel). As evident in this residual map, bright off-axis sources exhibit prominent radial calibration artifacts that direction-independent self-calibration cannot remove. We therefore used the PyBDSF source model (Fig.~\ref{fig:pybdsf}, middle panel) to construct a clean mask and identify the bright off-axis sources required for the subsequent direction-dependent calibration stage.

The final (third generation) calibration accounts for direction-dependent effects. After reviewing the FITS images generated from self-calibration, significant artifacting was noticed surrounding specific, off image-center bright sources, regardless of the weighting. Third generation calibration aims to correct these errors caused by ionospheric shifts and primary beam patterns through partitioning the image into a grid, calculating Jones matrices for each given direction of a bright source, and finally applying the corrections while reconstructing the image.

From the full 32k-channel L-band observation (spanning 856--1712 MHz with $26.123\text{ kHz}$ channel width), we select the contiguous, RFI-quiet frequency window spanning 1313--1508 MHz for our science analysis, represented as the unshaded region in Fig.~\ref{fig:full_rms}. This selected sub-band comprises $\sim 7500$ channels with the lowest and most stable noise floor. To reduce the computational cost of direction-dependent calibration and widefield deconvolution, the calibrated visibilities within the selected 1313--1508 MHz band were spectrally averaged by a factor of 10, yielding 750 channels with a channel width of $261.2\text{ kHz}$. The expected synchrotron emission from WIMP annihilation produces a smooth, broadband continuum, hence we believe this channel binning preserves the full spectral profile of any potential DM signal.

\begin{figure*}[htbp]
        \centering
        \includegraphics[width=\textwidth]{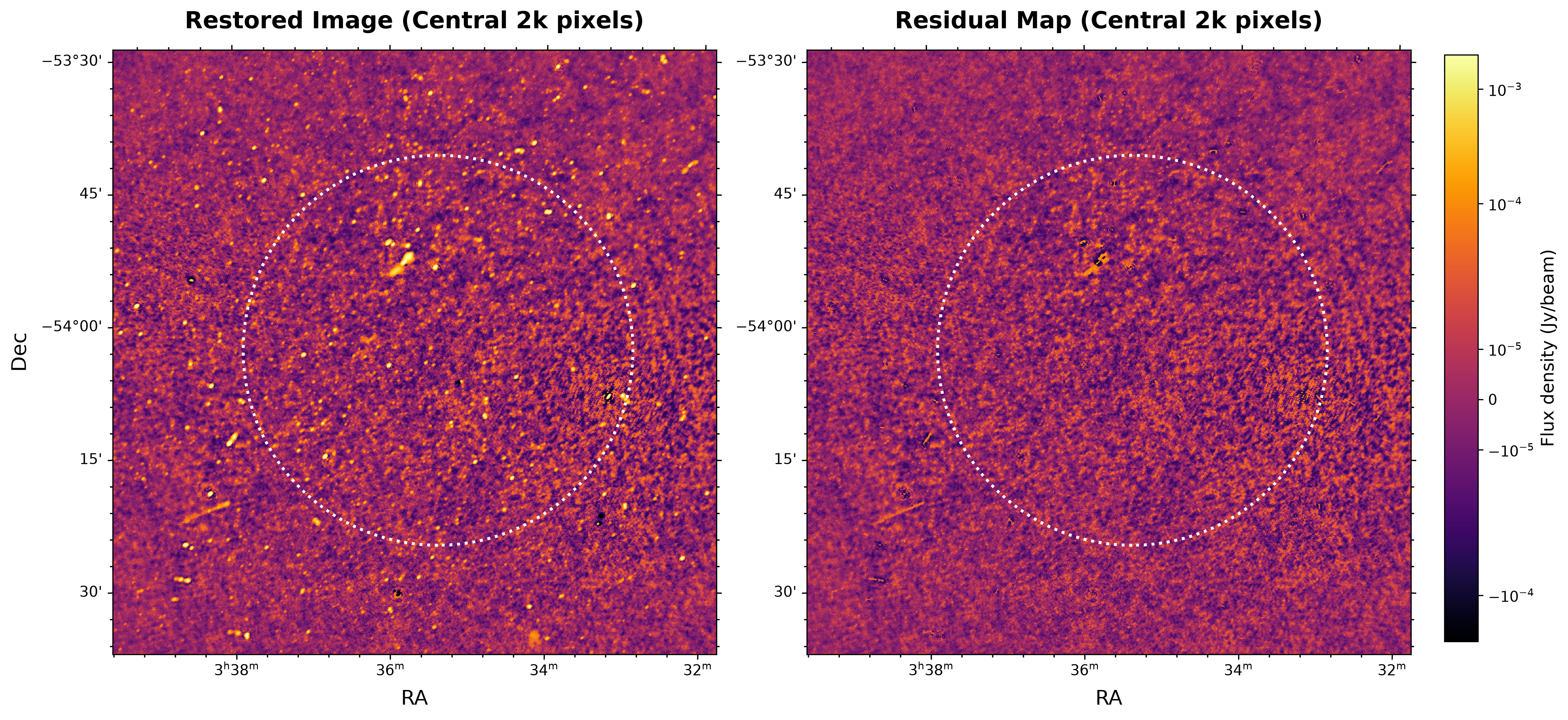} 
        \caption{The final inner $2048\times 2048$ pixel restored image of RetII (\textit{left} panel) and its corresponding residual map (\textit{right} panel). The scale radius of the DM halo $(\theta=22')$ is shown as the dashed white circle. As can be seen, the direction-dependent effects have been effectively calibrated, however the resolution of the image is weakened from the Briggs weighting choice.}
        \label{fig:full_subtraction}
    \end{figure*}

Final widefield MFS deconvolution was conducted across a large $6144 \times 6144$ (6k) pixel grid ($2''$ pixel scale) to deconvolve off-axis sources and sidelobes across the full primary beam. In the outer regions of this widefield map ($\theta \gtrsim 40'$ to $100'$ from the phase center), the observation achieves an instrumental thermal noise floor of $4.8\ \mu\text{Jy beam}^{-1}$. 
For our DM search, we extract the central $2048 \times 2048$ (2k) pixel cutout centered on RetII ($\theta \lesssim 34'$), shown in Fig.~\ref{fig:full_subtraction}. The science analysis directly utilizes the central residual map, which attains an RMS noise of $12.6\ \mu\text{Jy beam}^{-1}$. This noise is elevated relative to the outer widefield map due to faint, unmodeled background sources below the deconvolution threshold that are received at peak primary beam gain at the phase center, whereas in the outer field, source emission is substantially suppressed by primary beam attenuation. Furthermore, because the 0.5 Briggs-weighted synthesized beam ($\approx 7.5'' \times 7.0''$) is substantially oversampled by the $2''$ pixel grid, pixel noise in the residual map is inherently spatially correlated on the scale of the beam, an effect that is rigorously accounted for in our covariance analysis (Section~\ref{sec:likelihood}).

\section{Theoretical model} \label{sec:model}

In this section, we compute the synchrotron emission expected from WIMP annihilation and decay. We first describe the dark matter halo profile and then introduce three magnetic field models for the dwarf galaxy. Combining these ingredients, we calculate the emissivity of the relativistic electrons and positrons produced by dark matter annihilation and decay, and derive the resulting synchrotron flux (see also \cite{Fornengo_2012}).

\subsection{Dark Matter Density Profile} 
\label{sec:DM_density}

For the synchrotron emission from dark matter annihilation or decay, the spatial distribution of the dark matter halo is the starting point. For dwarf galaxies, observations and simulations indicate that the Einasto profile provides a better description of the mass density than the commonly used Navarro-Frenk-White (NFW) profile \cite{Chemin2011, DelPopolo:2012gd}. We therefore adopt the Einasto profile as our fiducial halo model:

\begin{equation}
    \rho(r) = \rho_{\rm s}\exp \left( -\frac{2}{\alpha_{\rm Ein}} \left[\left( \frac{r}{r_{\rm s}} \right)^{\alpha_{\rm Ein}} -1 \right] \right),
    \label{eq:dmdens}
\end{equation}
in which we take $\alpha_{\rm Ein}=0.4$ as the controlling parameter for the steepness of the slope, $\rho_{\rm s} =7.0 \times 10^{7}\, \Msun\, {\rm kpc}^{-3}$ for the characteristic density and $r_{\rm s}=0.2$ kpc for the scale radius~\cite{Bonnivard_2015, Regis_2017}. The enclosed mass within the half-light radius in Table~\ref{tab:RetII} was obtained from the spectroscopic observations of \citeauthor{Simon_2015} \cite{Simon_2015}. We have checked that the integrated mass within $r_{\rm 1/2}=55$ pc for this Einasto profile is consistent with this observation.

\subsection{Magnetic field models}
The magnetic field properties of the target region play a central role in determining the synchrotron emission from dark matter annihilation or decay. In dwarf spheroidal galaxies, however, both the field strength and its spatial morphology are poorly constrained by observations, which introduces a significant uncertainty in the predicted synchrotron signal \cite{Regis_2017}. We consider three distinct spatial profiles for the magnetic field $B(r)$, each normalized to a characteristic central field strength $B_0 \equiv B(r=0)$.

\subsubsection{Uniform Magnetic field}
\label{sec:uniform_B}

There are two primary mechanisms that could source an ambient magnetic field in a dSph galaxy: an internal dynamo associated with local star formation, and magnetization inherited from the surrounding medium \cite{Chy_y_2011, Regis_2017}. In the case of ultra-faint dSphs like RetII, the stellar population is ancient with no active star formation for $\sim 10$ billion years \cite{2023ApJ...944...43S}; we therefore assume any internal dynamo is inactive and focus on the surrounding medium.

For the surrounding medium, studies of galaxy clusters and the Milky Way halo indicate pervasive magnetized environments. Studies of the Coma Cluster reveal Intracluster Medium (ICM) field strengths of $0.2$--$0.4\ \mu\text{G}$ extending out to $\sim 4\text{ Mpc}$ \cite{Kronberg_2007}. RetII orbits within the outer halo of the Milky Way ($D = 31.2\text{ kpc}$) at high Galactic latitude. From the Faraday Rotation Measures of extragalactic sources mapped across the Galactic halo by \citeauthor{Taylor_2009} \cite{Taylor_2009}, the coherent Galactic magnetic field is approximately $0.83\ \mu\text{G}$. More recently, comprehensive modeling of WMAP and Planck data \cite{Unger_2024} found that the coherent field strength in the Galactic halo is $\mathcal{O}(1)\ \mu\text{G}$, motivating our choice of $B_{0}=1\ \mu\text{G}$. Assuming this field remains completely uniform ($B(r) = B_0$) and unattenuated across the entire DM halo represents an optimistic scenario, maximizing the synchrotron emissivity of diffusing $e^\pm$ pairs and yielding our most stringent exclusion limits.

\subsubsection{Self-confined Magnetic field}

For this intermediate approach, we consider magnetic turbulence that is self-generated by the DM-induced $e^{\pm}$. In star-forming galaxies, magnetic fluctuations driven by cosmic-ray streaming instabilities are typically subdominant compared to the vigorous turbulence driven by stellar feedback. In a system like RetII with no ongoing star formation, however, self-generated magnetic turbulence can become an important effect. The irregularities and fluctuations in the ambient field induced by the injected $e^{\pm}$ particles enhance the local turbulence, significantly prolonging the residence time of $e^{\pm}$ in the central emission region \cite{Regis_2023}. 

Interestingly, this self-confinement mechanism creates a competing effect between the ambient field strength and the synchrotron output. As noted by \citeauthor{Semane_2026} \cite{Semane_2026}, a lower baseline field $B_0$ reduces the raw synchrotron emissivity but simultaneously enhances particle confinement in the core, causing the predicted emission for heavier WIMPs to be relatively insensitive to variations in $B_0$. We adopt a standard beta-profile for the self-confined field:
\begin{equation}
    B(r) = B_0 \left[ 1 + \left(\frac{r}{r_{\rm s}}\right)^2 \right]^{-\beta},
\end{equation}
with $B_{0}=1\,\mu\mathrm{G}$ chosen to be consistent with the uniform profile and $\beta=0.5$ being a consequence of the magnetic energy density tracing the DM mass density for the self-confined regime. Here $r_{\rm s}$ is also the scale radius as in the Einasto profile.

\subsubsection{Exponentially Decaying Magnetic field}

We now consider a highly pessimistic scenario, for which the magnetic field strength is modeled as an exponential profile:
\begin{equation}
    B(r)=B_{0} \exp \bigg( -\frac{r}{r_{\rm 1/2}} \bigg),
\end{equation}
using the characteristic decay length set to the stellar half-light radius of the galaxy, $r_{\rm 1/2}=55$ pc, and where we set $B_{0}=1$ $\mu$G for consistency. This exponential profile represents an intrinsic relic field scenario, where the remnant magnetic field is localized to the stellar core. The magnetic energy density scales as $U_B \propto \exp(-2r/r_{1/2})$, and as such the field is negligible outside of the stellar core. The expected signal generated from this particular profile will be drastically diminished due to the confinement region being severely limited in comparison with the diffusion length. Since the synchrotron emission scales with $B^{2}$, observable emission is spatially restricted to $r \lesssim r_{1/2}$, rather than originating from the full DM halo. It has been noted that the injected $e^{\pm}$ plasma in the extremely weak magnetic field will cause plasma instabilities, thereby amplifying the local field, which will revert this profile to the self-confinement regime \cite{Regis_2023}. 

As such, while we consider the self-confinement case to be the bounding floor for the physically realizable exclusion limits, we also consider the exponentially decaying profile to represent the extreme limit in which self-confinement mechanisms fail to operate efficiently. This could be due to an exceptionally low interaction cross-section, or in scenarios dominated by slow DM decay.

\subsection{$e^{\pm}$ Diffusion and the Flux Density}

In order to calculate the synchrotron emission from dark matter annihilation or decay, we solve the diffusion equation under the steady-state assumption to obtain the equilibrium $e^{\pm}$ spectrum. Here, we assume spherical symmetry and consider \cite{Beck_2023, McDaniel_2017}
\begin{equation}
    \begin{split}
        \frac{\partial}{\partial t}\psi_{e} =& \nabla \big(D(E,r)\nabla\psi_{e} \big) + \frac{\partial}{\partial E}\big( b(E,r)\psi_{e}\big) \\
        &+ Q_{e}(E,r),
    \end{split}
\end{equation}
where $\psi_{e}\equiv \partial n_{\rm e}/\partial E$ is the equilibrium $e^{\pm}$ spectrum, $D(E,r)$ is the diffusion coefficient, $b(E,r)$ is the energy-loss term, and $Q_{e}(E,r)$ is the electron source term. We assume that the source term $Q_{e}(E,r)$ is entirely determined by dark matter and is given, in the case of WIMP annihilation, by
\begin{equation}
    Q_{e}(r,E) = \frac{1}{2} \langle\sigma v \rangle 
    \left(\frac{\rho_{\chi}(r)}{m_{\chi}}\right)^{2}
    \sum_{f}\frac{\text{d} N^{f}_{e}}{\text{d}E} ,
    \label{eq:annhsource}
\end{equation}
where $r$ is used as the distance from the center of RetII, $E$ the energy of the injected $e^{\pm}$, $\langle \sigma v \rangle$ the velocity averaged annihilation cross section, $m_{\chi}$ the WIMP mass under consideration, and $\rho_{\chi}(r)$ the DM density profile. In the case of WIMP decay
\begin{equation}
    Q_{e}(r,E) = \Gamma \left( \frac{\rho_{\chi}(r)}{m_{\chi}}\right) \sum_{f}\frac{\text{d} N^{f}_{e}}{\text{d}E} ,
    \label{eq:decaysource}
\end{equation}
where $\Gamma$ is the decay rate. The expression d$N^{f}_{e}/\text{d}E$ appearing in both Eqs.~\ref{eq:annhsource} and~\ref{eq:decaysource} represents the $e^{\pm}$ energy spectrum originating from the DM source which annihilates/decays into the final state. These spectra are obtained from pre-computed numerical tables provided by the Poor Particle Physicist Cookbook for Dark Matter Indirect Detection (PPPC4DMID~\cite{Cirelli:2010xx}). The diffusion coefficient is modeled through
\begin{equation}
    D(E,r)=D_{0} \bigg( \frac{E}{\text{1 GeV}} \bigg)^{\delta} \bigg( \frac{B(r)}{B(0)} \bigg)^{-\delta},
\end{equation}
where we adopt $\delta=0.3$, and $B(r)$ denotes the spatially varying magnetic-field strength as the Kolmogorov turbulence for interstellar fluctuations~\cite{Colafrancesco:2006he, sarkis2024darkmatterspowerfultoolwimpy}. A note is to be made here regarding the diffusion coefficient $D_{0}$. It can conservatively be assumed that this value is of a similar order of magnitude to that of the Milky Way, where $D_{0}=3.1\times10^{28} \text{ cm}^{2}\text{ s}^{-1}$ \cite{hu2024constraintsannihilationheavydark}, while \citeauthor{Gajovi__2023} \cite{Gajovi__2023} opted for a value of $D_{0}=3.1\times10^{27} \text{ cm}^{2}\text{ s}^{-1}$ and varied their results by two orders of magnitude, or even as low as $D_{0}=3\times10^{26} \text{ cm}^{2}\text{s}^{-1}$ from \citeauthor{Kar:2019hnj} \cite{Kar:2019hnj}. The exact scaling of this coefficient is uncertain for dSphs, and is discussed further in Section~\ref{sec:results_discussion}. We set the diffusion normalization parameter to be uniform, fixed at $D_{0}=3.1 \times 10^{28} \text{ cm}^{2}\text{ s}^{-1}$. We make this assumption as a somewhat conservative approach, since the coherent magnetic field from the halo component of the Milky Way shows evidence for extending several kpc from the center \cite{Unger_2024}. The full energy loss equation is then described through \cite{sarkis2024darkmatterspowerfultoolwimpy}

\begin{equation}
    \begin{split}
        b(E,r) =&  \left[ b_{\text{IC}} + b_{\text{syn}} \left(\frac{B(r)}{\text{1 }\mu\text{G}}\right)^{2} \right] \left(\frac{E}{\text{1 GeV}}\right)^{2} \\
        &+ b_{\text{Coul}} \bigg( \frac{n_{\rm e}(r)}{\text{1 cm}^{-3}} \bigg)\Bigg[\log \Big(\frac{\gamma}{n_{\rm e}(r) / 1\,{\rm cm^{-3}}}\Big) +73 \Bigg] \\ 
        &+ b_{\rm brem}\Big(\frac{n_{\rm e}(r)}{1\,{\rm cm^{-3}}}\Big)\gamma\big(\log(\gamma) + 0.36\big),
    \end{split}
\end{equation}
where $\gamma$ is the usual Lorentz factor of the electron.

The energy-loss coefficients are in units of $\rm GeV\,s^{-1}$ and given by
\begin{equation}
\begin{aligned}
   & b_{\rm IC}  = 0.26\times10^{-16},\\
   & b_{\rm syn}  =0.025\times10^{-16},\\
   & b_{\rm Coul} =7.6\times10^{-18},\\
   & b_{\rm brem}  =7.1\times10^{-20},
\end{aligned}
\end{equation}
corresponding to inverse Compton scattering, synchrotron radiation, Coulomb interactions, and bremsstrahlung, respectively. Given the low gas content of dSphs, we neglect the Coulomb and bremsstrahlung energy-loss terms. The dominant contribution to the energy losses therefore comes from inverse Compton scattering off the CMB. For the magnetic-field strengths considered here, $B\lesssim1\,\mu\mathrm{G}$, synchrotron losses provide only a subdominant contribution to the total energy-loss rate. Nevertheless, synchrotron emission provides an important observational channel for probing the $e^\pm$ produced by dark matter annihilation or decay, and its relatively weak contribution to the energy losses in dSphs leads to a correspondingly suppressed synchrotron signal.

\begin{figure*}[htbp]
    \centering
    \includegraphics[width=0.8\textwidth]{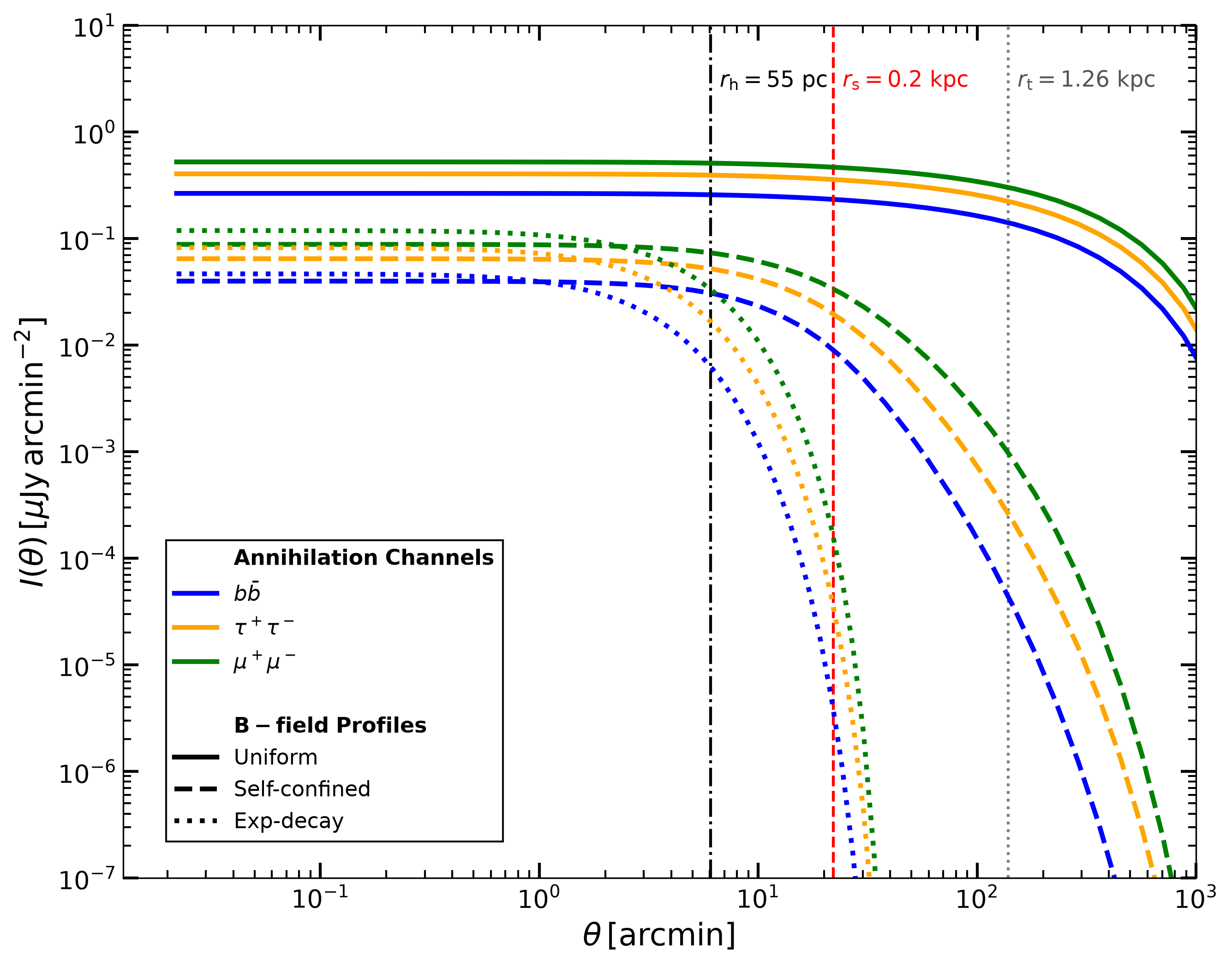}
    \caption{Impact of different magnetic field models (uniform, self-confined and exponentially decaying) with the same central strength $B_{0}=1\,\mu{\rm G}$ and different decay channels ($b\bar{b}$, $\tau^{+}\tau^{-}$ and $\mu^{+}\mu^{-}$) on the surface brightness profile $I(\theta)$ for a given dark matter mass $m_{\chi}=100\,{\rm GeV}$, frequency $\nu=1400\,{\rm MHz}$, cross-section $\langle \sigma v \rangle = 10^{-26} \text{cm}^3\,\text{ s}^{-1}$, diffuse constant $D_{0}=3.1\times 10^{28}\text{ cm}^2\text{ s}^{-1}$. The black dash-dotted, red dashed and grey dotted vertical lines indicate the locations of the half-light radius ($r_{\rm h}=55\,{\rm pc}$), the scale radius ($r_{\rm s}=0.2\,{\rm kpc}$) and the tidal radius ($r_{\rm t}=1.26\,{\rm kpc}$) respectively. Note that although all magnetic field models are normalized to the same central value, the central surface brightness is higher for the uniform profile as the surface brightness integrates emissivity along the full line-of-sight column through the halo, whereas the self-confined and exponentially decaying profiles attenuate outside their characteristic scale radii.}
    \label{fig:surf_comb}
\end{figure*}

The synchrotron emissivity at frequency $\nu$ and radius $r$ is given by
\begin{equation}
    j_{\text{syn}}(\nu,r) = \int^{m_{\chi}}_{m_{\rm e}}{\rm d}E \hspace{0.2em} \Psi(E,r) \hspace{0.2em} P_{\text{syn}}(\nu, E, B),
\end{equation}
where $P_{\rm syn}(\nu,E,B)$ denotes the synchrotron power emitted at frequency $\nu$ by an $e^\pm$ of energy $E$ in a magnetic field of strength $B$ and $\Psi(E,r)=(\psi_{e^-}(E,r)+\psi_{e^{+}}(E,r))$ is the contribution of both positrons and electrons. The corresponding synchrotron surface brightness is obtained by integrating the emissivity along the line of sight. If we assume spherical symmetry, then the isometric surface brightness can be written as

\begin{equation}
    I_{\text{syn}}(\nu, \theta) = \frac{1}{2\pi}\int^{\infty}_{0} {\rm d}l \hspace{0.2em} j_{\text{syn}}\bigg(\nu, \sqrt{(D\hspace{0.1em}\theta)^{2}+l^{2}} \bigg),
\end{equation}
where $l$ denotes the line-of-sight coordinate and $R=D\theta$ is the projected radius at an angular separation $\theta$ from the center of the target, with $D$ denoting its distance. Here, we have used the small-angle approximation, $R\simeq D\theta$, which is valid for $R\ll D$. In this work, we employ the publicly available \texttt{DarkMatters} package \cite{sarkis2024darkmatterspowerfultoolwimpy} to perform the calculations. Annihilation emission for 100 GeV mass WIMPs using $b\bar{b}$, $\tau^{+}\tau^{-}$ and $\mu^{+}\mu^{-}$ channels can be seen in Fig.~\ref{fig:surf_comb}. It is to be noted that the example frequency for these models is set at $\sim$1400 MHz, chosen as the midpoint frequency channel of our sample. The predicted intensities are highly dependent on the specific magnetic field profile under consideration. As can be seen, while each profile has the same normalization ($B_{0}=1\,\mu\text{G}$), the uniform profile has a considerably larger intensity, due to the synchrotron emissivity being integrated along the line-of-sight through the halo. For the uniform case, $B_{0}$ is maintained throughout the entire line-of-sight column, whereas for the self-confined and exponentially decaying profiles, $B(r)$ drops rapidly. As discussed, the self-confined case follows a beta-profile for the magnetic field, hence the initial intensity is lower than the exponential case. However, its magnetic field also extends further out, resulting in the exponential surface brightness profile dropping off much faster.

\section{Statistical Likelihood}
\label{sec:likelihood}

To constrain WIMP DM properties, we perform hypothesis testing by comparing the observed radio map against predicted synchrotron templates using a multivariate Gaussian likelihood. As the DM halo and resulting synchrotron emissivity are spherically symmetric, we compress the 2D residual map into a 1D data vector of azimuthally-averaged surface brightness bins ($I_{\rm obs}(\theta)$), centered on RetII with a radial bin width of $\Delta\theta = 2.5'$. This compression preserves the full radial profile of any potential DM signal while averaging out uncorrelated azimuthal noise.

Constructing our likelihood requires an accurate covariance matrix $\mathbf{C}$ that captures the spatial correlations of the residual background. We assume that the residual noise field is well-approximated as a Gaussian Random Field (GRF). Inherent to interferometric imaging, short-range spatial correlations exist between adjacent pixels due to convolution with the synthesized beam. In addition, longer-range spatial correlations may be present as a result of faint residual background sources below the deconvolution threshold and low-level direction-dependent calibration residuals. Across the central science cutout, we assume that these statistical properties are statistically stationary (homogeneous and isotropic).

Under the assumption of statistical stationarity, the two-point spatial correlations of the image can be characterized either in real space via the 2D auto-correlation function (ACF) or, equivalently, in Fourier space via the spherically-averaged spatial power spectrum $P(k)$ (related by the Wiener–Khinchin theorem). The central dwarf core ($r_{1/2} = 6.1'$) occupies only $\sim 2\%$ of the total $2048 \times 2048$ Fourier domain, hence any potential DM contribution to the global power spectrum is negligible compared to the pervasive instrumental and beam noise. 

\begin{figure*}[htbp]
    \centering
\centerline{\includegraphics[width=0.55\textwidth]{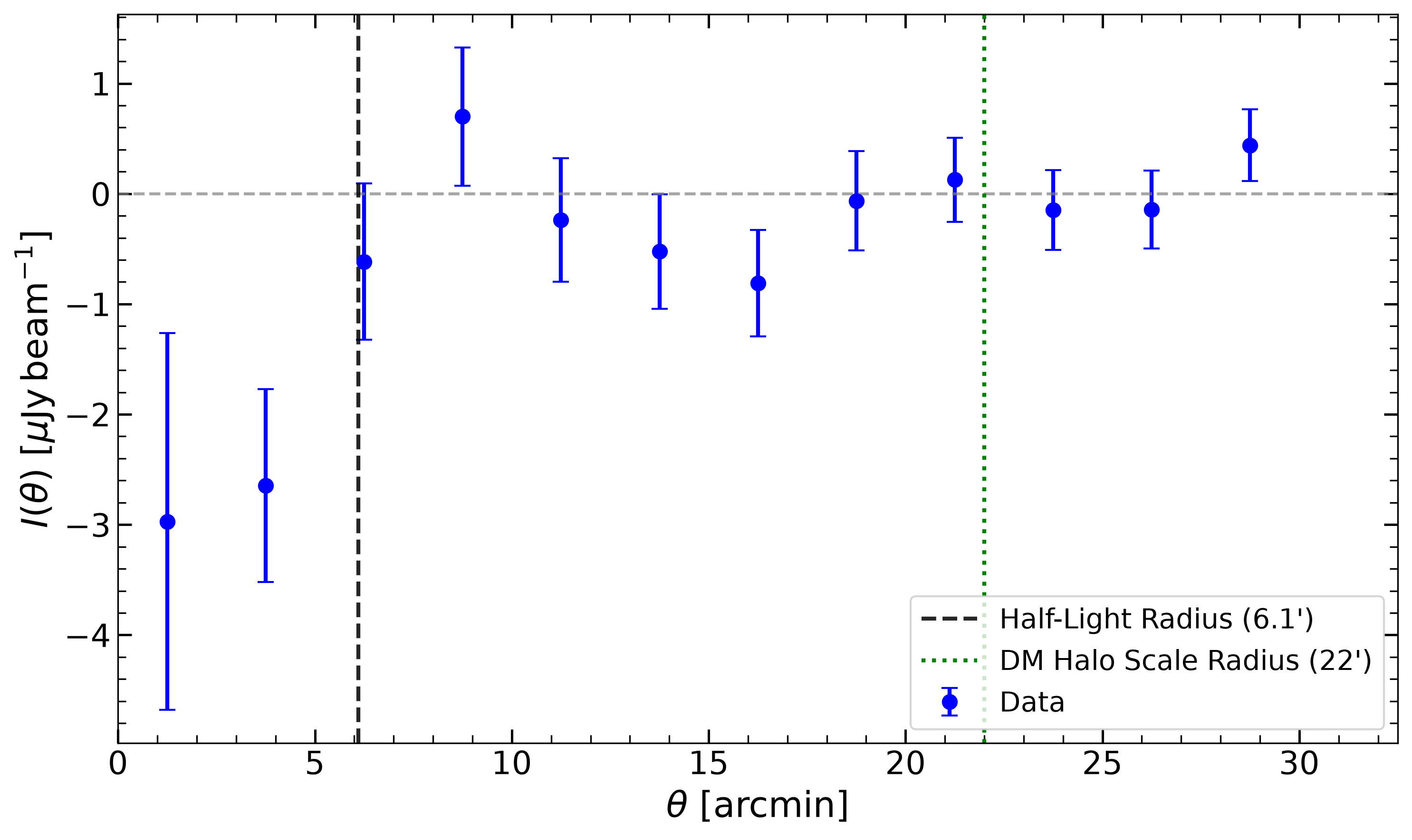}
\includegraphics[width=0.45\textwidth]{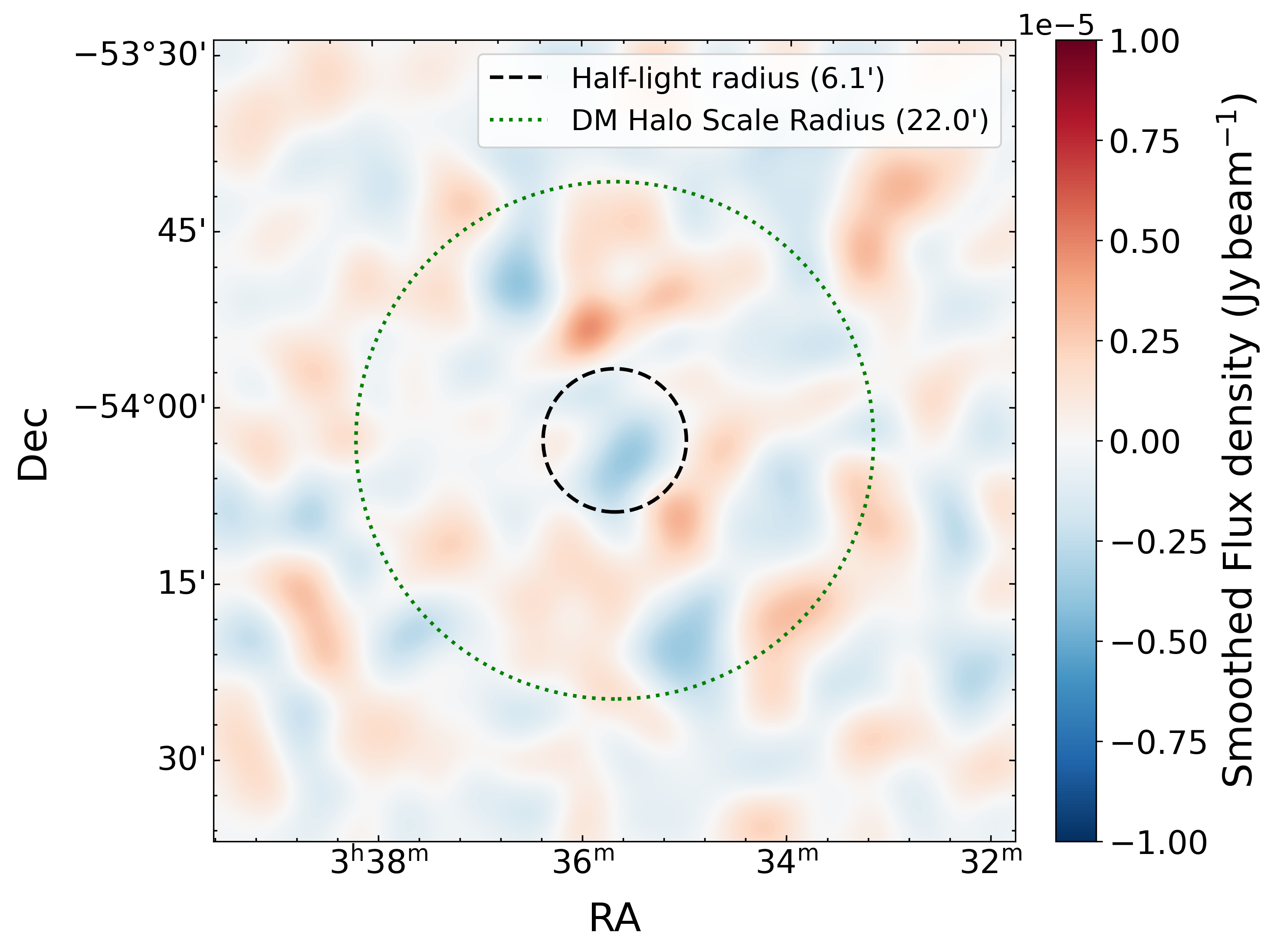}}
    \caption{\textit{Left}--The radial brightness profile of the residual image derived through the power spectrum method. The $\pm 1\sigma$ errors are determined through the variance between the respective bin of each profile across all realizations of the phase of the noise, where we chose $N=1000$ samples to generate the profile.
    \textit{Right}--The L-band residual image of Reticulum II, smoothed with a Gaussian kernel to highlight the large-scale negative flux systematic in the core. In both panels, the inner black dashed circle/line marks the half-light radius $\theta=6.1'$, and the outer green dotted circle/line denotes the dark matter halo scale radius $\theta=22.0'$. 
    }
    \label{fig:residual_analysis}
\end{figure*}

\begin{figure}[htbp]
        \centering
        \includegraphics[width=0.48\textwidth]{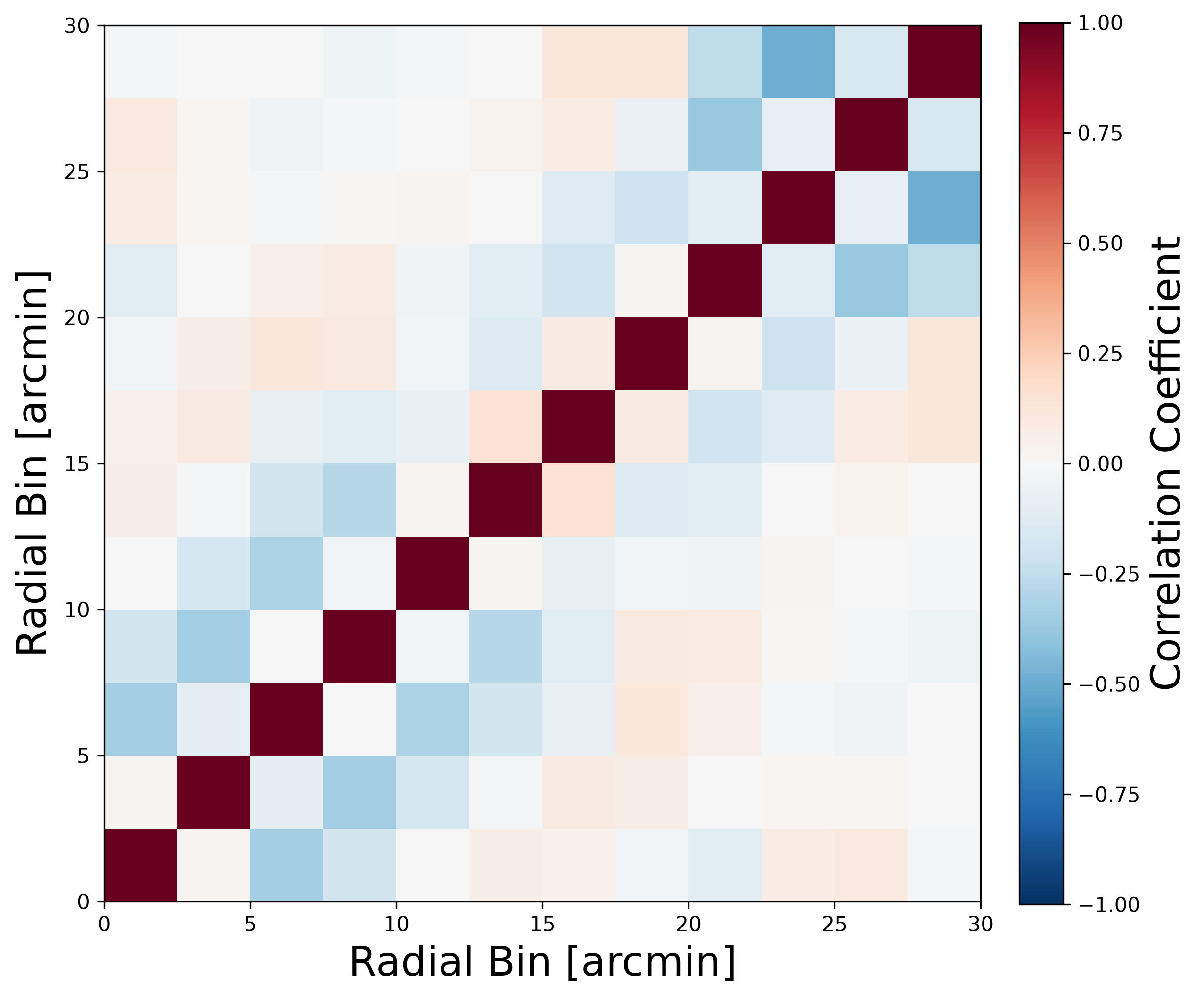}  
        \caption{Spatial correlation matrix of the surface brightness background, derived from $N=1000$ simulated profiles of the residual map.}
        \label{fig:corr}
\end{figure}

We measure the 1D isotropic power spectrum $P(k)$ directly from the 2D residual map using the \texttt{powerbox} package \cite{G_Murray_2018}. To determine the full covariance matrix between radial bins without relying on idealized beam models, we use \texttt{powerbox} to generate $N = 1000$ synthetic realizations of the 2D Gaussian random field. For each realization, the Fourier amplitudes are drawn with standard deviation proportional to $\sqrt{P(k)}$ and assigned uniform random phases $\phi(\mathbf{k}) \in [0, 2\pi)$, after which an inverse 2D FFT reconstructs a real-space mock residual map. For each mock map $k$, we extract the radial profile $I_{\rm mock}^k(\theta)$ using the same $2.5'$ annular binning. The sample variance across all realizations defines the empirical error bars shown in the left panel of Fig.~\ref{fig:residual_analysis}, and the covariance between annular bins $\theta_a$ and $\theta_b$ is computed as:
\begin{equation}\label{eq:covariance}
    C(\theta_{a}, \theta_{b}) = \frac{1}{N} \sum_{k=1}^{N} I_{\rm mock}^{k}(\theta_{a}) I_{\rm mock}^{k}(\theta_{b}),
\end{equation}
where $I_{\rm mock}^{k}(\theta)$ is the azimuthally-averaged intensity in the annular bin centered at $\theta$ for the $k$-th mock realization. As the mock Gaussian random fields have an expected mean of zero by construction ($\langle I_{\rm mock}(\theta) \rangle = 0$), the sample mean term is omitted. Normalizing this covariance matrix yields the spatial correlation matrix shown in Fig.~\ref{fig:corr}, which illustrates the non-negligible off-diagonal correlations that extend across adjacent radial bins.


In the left panel of Fig.~\ref{fig:residual_analysis}, the observed radial profile exhibits a negative surface brightness in the innermost annular bins ($I_{\rm obs} \approx -3\ \mu\text{Jy beam}^{-1}$). This feature is visible in the right panel of Fig.~\ref{fig:residual_analysis}, which displays the residual map smoothed with a Gaussian kernel matching the bin size ($2.5'$). Smoothing a noise field at this scale by construction results in apparent structures at the smoothing scale. However, note that the amplitude of the smoothed fluctuations is significantly lower than the per-pixel noise, indicating that while the noise may be correlated pixel-to-pixel, it is not fully correlated at these angular scales. In isolation, these smoothed structures represent a $\sim 1.5\text{--}2\sigma$ deviation, which is roughly consistent with random fluctuations of the noise field.

There are two aspects that call into question whether this central depression can be explained as pure random noise. Firstly, both innermost radial bins are systematically below zero, with the second bin possessing an even smaller variance and thus higher individual statistical significance ($\sim 3\sigma$). Secondly, the smoothed 2D residual map reveals a somewhat structured pattern: the central negative depression is accompanied by adjacent positive ``side-blobs''. Such alternating negative and positive structures are characteristic of mild interferometric calibration or deconvolution artifacts, most likely originating from residual sidelobes of imperfectly cleaned off-axis sources. While this systematic effect is relatively modest, it is verging on statistically significant and coincides directly with the core of RetII.

However, DM annihilation predicts a strictly positive, centrally peaked emission profile ($I_{\rm syn}(\theta) \ge 0$); this localized negative fluctuation presents a challenge for limit setting. An unadjusted likelihood evaluated against a negative central trough would penalize positive signal models, artificially driving the upper limits on $\langle\sigma v\rangle$ to be overly optimistic. Modeling this feature with an explicit additive background template is intractable, as any central background component is almost perfectly degenerate with the sharply peaked DM profile, which we discuss in more detail in Sec.~\ref{sec:results_discussion}. 

To account for the slower averaging of correlated noise and to ensure that negative fluctuations do not artificially bias our constraints, we adopt an empirical approach \cite{Ma_2014}. We introduce a global variance scaling hyper-parameter $\alpha$ that inflates the baseline covariance matrix as $\alpha^2 \mathbf{C}$, allowing the total noise level to be boosted to absorb residual large-scale fluctuations in a model-independent manner. However, by uniformly scaling all bins, this approach unnecessarily penalizes the well-behaved outer annuli, yielding overly conservative limits, particularly for the self-confined and exponentially decaying profiles.

\begin{figure*}[htbp]
    \centering
    \includegraphics[width=\textwidth]{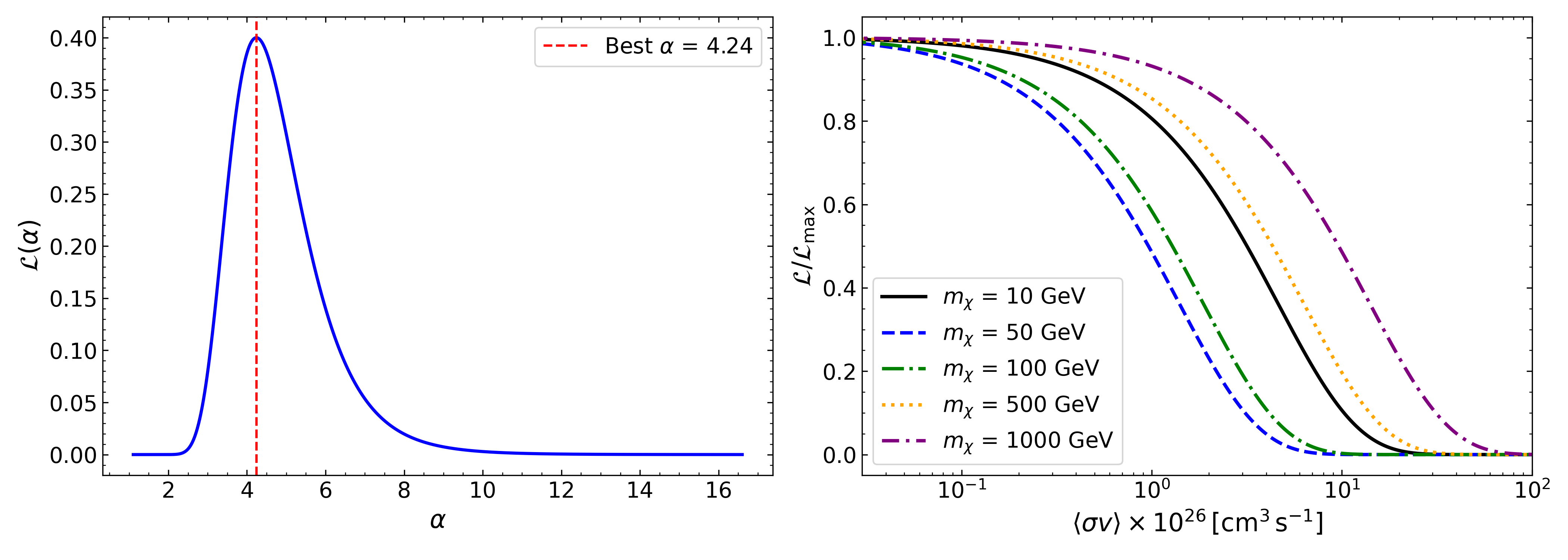} 
    \caption{\textit{Left}--The distribution of $\alpha$ across the effective likelihoods of the cross-section $\langle \sigma v\rangle$ and mass $m_{\chi}$. The vertical line shows the maximum likelihood $\alpha$ value, which is $4.24$. 
    \textit{Right}--The normalized likelihoods for a range of WIMP masses while setting $\alpha=4.24$.}
    \label{fig:alpha_value}
\end{figure*}

The $\chi^2$ test statistic is then constructed as
\begin{equation}
\begin{aligned}
    \chi^2 =  \sum_{a,b}
\Big(&I_{\rm obs}(\theta_a) - N_{\rm DM} I_{\rm syn}(\theta_a)
\Big)
\Big[\alpha^2 C(\theta_a,\theta_b)\Big]^{-1} \\
&\times
\Big(I_{\rm obs}(\theta_b) - N_{\rm DM} I_{\rm syn}(\theta_b)\Big),
\end{aligned}
\end{equation}
where $N_{\rm DM} = \langle\sigma v\rangle / \langle\sigma v\rangle_0$ is the ratio of the cross-section to a reference benchmark $\langle\sigma v\rangle_0 = 10^{-26}\,{\rm cm^{3}\,s^{-1}}$, $I_{\rm syn}(\theta)$ is the predicted synchrotron profile, and $C(\theta_a, \theta_b)$ is the covariance matrix computed from Eq.~\ref{eq:covariance}.
The multivariate Gaussian likelihood is given by
\begin{equation}
    \mathcal{L}(\langle \sigma v\rangle, m_{\chi},\alpha)=\frac{1}{\sqrt{\det(2\pi \alpha^2\mathbf{C})}}\exp\left[-\frac{\chi^2(N_{\rm DM}, m_{\chi}, \alpha)}{2}\right].
    \label{eq:likelihood}
\end{equation}
We determine the optimal noise inflation factor by maximizing the marginal likelihood over the remaining parameters:
\begin{equation}
    \mathcal{L}(\alpha) = \int_{0}^{\infty} {\rm d}N_{\rm DM} \int_{m_{\chi, \rm \min}}^{m_{\chi, \rm \max}} {\rm d}m_{\chi} \, \mathcal{L}(N_{\rm DM}, m_{\chi},\alpha),
    \label{eq:marginal_alpha}
\end{equation}
where flat priors are assumed for $\langle \sigma v\rangle$ and $m_{\chi}$ over their respective ranges ($m_{\chi} \in [10, 10^3]\text{ GeV}$). This yields a best-fit value of $\alpha = 4.24$, as shown in Fig.~\ref{fig:alpha_value}. We fix $\alpha = 4.24$ in the subsequent analysis, scaling the baseline covariance by $\alpha^2 \approx 18$.

While no individual radial bin deviates from zero at a level that would independently require an inflation factor as large as $\alpha= 4.24$, the likelihood operates on the full off-diagonal covariance matrix $\mathbf{C}$. Adjacent bins exhibit positive spatial correlations, so coherent offsets in the same direction are statistically less favored under the multivariate likelihood than independent variances. Bringing these correlated offsets simultaneously into consistency with zero therefore requires a larger overall variance inflation than a naive bin-by-bin assessment would suggest.

Finally, the 95\% confidence level upper limit cross-section $\langle \sigma v\rangle_{95}$ for each fixed WIMP mass $m_\chi$ and fixed $\alpha = 4.24$ is determined using the one-sided profile likelihood ratio test statistic:
\begin{equation}
    \Delta\chi^2 = \chi^2(\langle\sigma v\rangle_{95}, m_\chi, \alpha) - \chi^2(\langle\sigma v\rangle_{\rm best}, m_\chi, \alpha),
    \label{eq:upper_limit_fixed_m}
\end{equation}
where $\langle\sigma v\rangle_{\rm best}$ is the best-fit cross-section constrained to the physical region $\langle\sigma v\rangle \ge 0$, and $\Delta\chi^2 = 2.71$ corresponds to the 95\% one-sided confidence threshold.

\section{Results and Discussion} \label{sec:results_discussion}

\begin{figure*}[htbp]
    \begin{subfigure}{0.75\textwidth}
        \includegraphics[width=\textwidth]{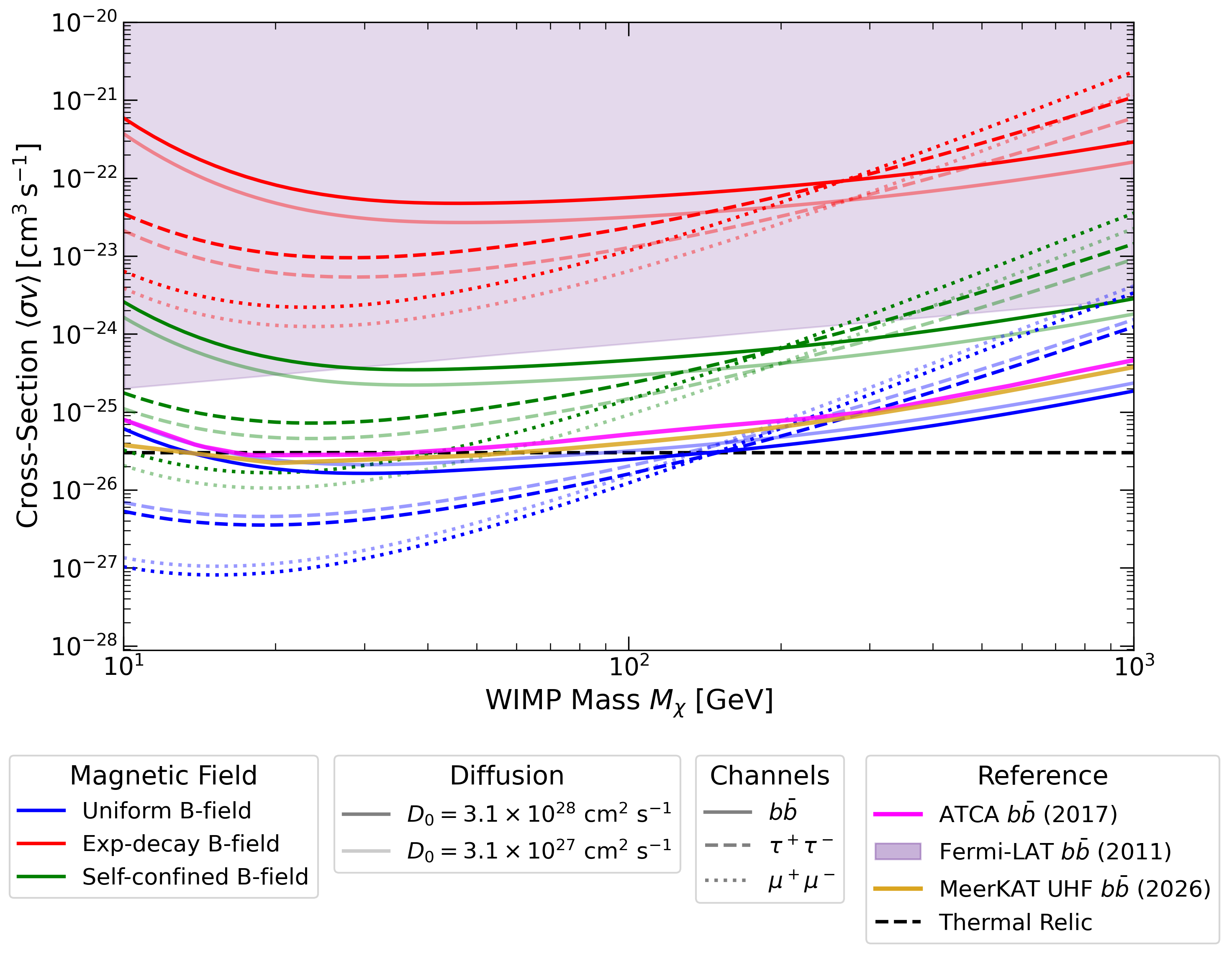}  
    \end{subfigure}
    \hfill
    \begin{subfigure}{0.75\textwidth}
        \centering
        \includegraphics[width=\textwidth]{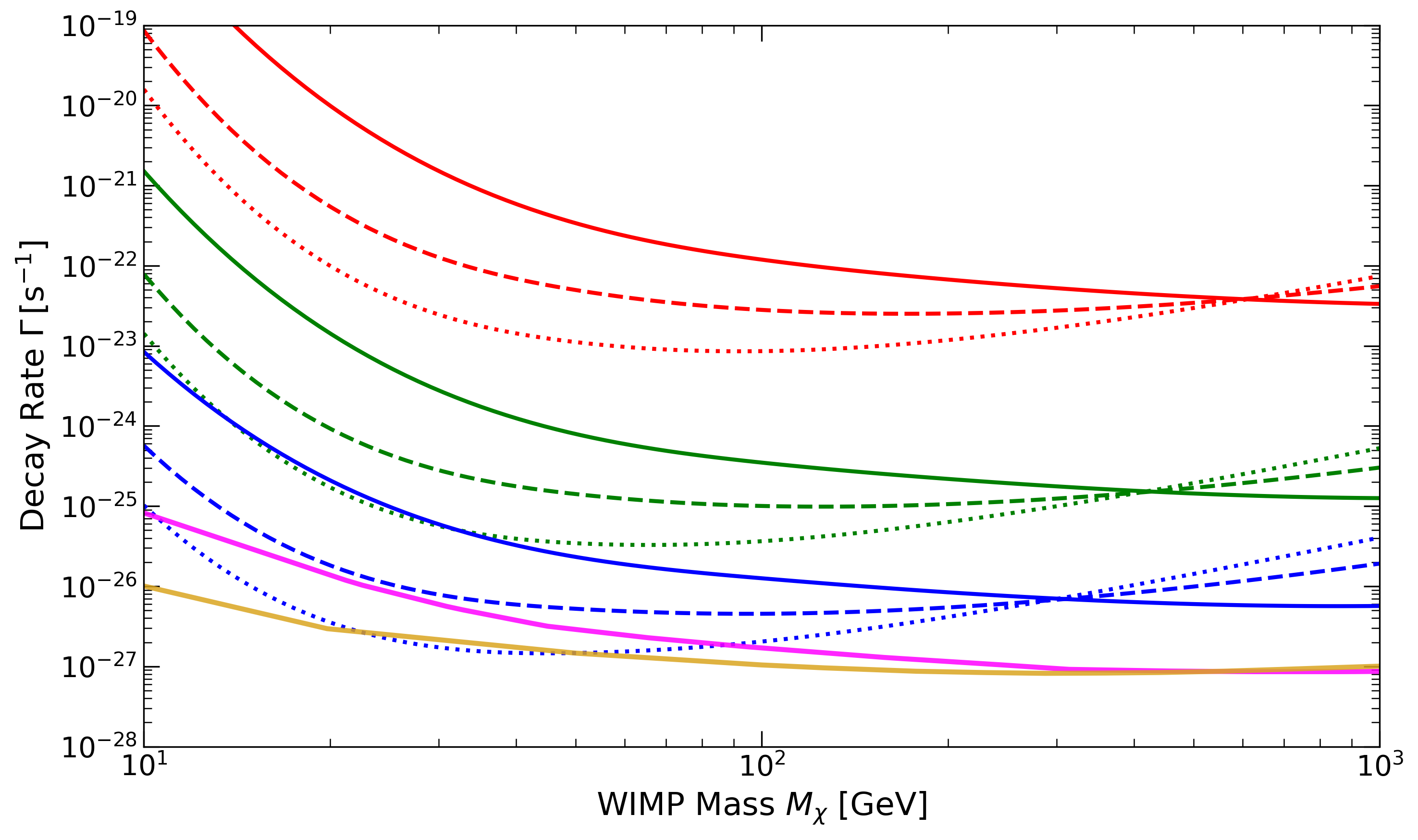} 
    \end{subfigure}
    \caption{The 95\% C.L. exclusion limits for WIMP masses using $b\bar{b}$, $\tau^{+}\tau^{-}$ and $\mu^{+}\mu^{-}$ annihilation ({\it Top}, which also includes a comparison between the diffusion coefficients) and decay ({\it Bottom}, only using $D_0=3.1\times 10^{28}\text{ cm}^2\text{ s}^{-1}$) production mechanisms. We also include references to previous limits derived from MeerKAT UHF \cite{Semane_2026}, ATCA \cite{Regis_2017} and specifically for the cross-section limits, Fermi-LAT \cite{Donato_2012} is shown as a shaded region to represent the most stringent limits. For each of these external references, only the benchmark $b\bar{b}$ channel is shown for simplicity.}
    \label{fig:excl}
\end{figure*}

The upper panel of Fig.~\ref{fig:excl} shows the 95\% confidence upper limits on the WIMP annihilation cross-section for WIMP masses between 10 GeV and 1 TeV. The limits are derived for three annihilation channels, $b\bar{b}$, $\mu^{+}\mu^{-}$, and $\tau^{+}\tau^{-}$, three magnetic-field models, and different values of the diffusion coefficient $D_0$, yielding 18 exclusion curves in total. Since the signal arises from synchrotron emission by WIMP-induced electrons and positrons, the resulting limits depend on both the assumed magnetic-field morphology and the efficiency of electron diffusion.

For $D_0 = 3.1\times 10^{28}\ \mathrm{cm^2\,s^{-1}}$, the strongest limits are obtained for the uniform magnetic-field model. In this case, the high-energy electrons and positrons produce synchrotron emission efficiently throughout the region where they propagate, so the predicted synchrotron signal is maximal and the resulting constraint is the most stringent. For the exponentially decaying magnetic field, the field strength is concentrated near the centre of the dwarf galaxy and drops rapidly with radius. Since synchrotron emission depends strongly on the magnetic-field strength, the dominant contribution comes from the central region and the signal falls off quickly outward, leading to weaker limits than in the uniform case. The self-confined magnetic field model yields limits intermediate between these two cases, as expected from its intermediate spatial dependence.

When the diffusion coefficient is reduced to $D_0 = 3.1\times 10^{27}\, \mathrm{cm^2\,s^{-1}}$, high-energy electrons and positrons remain confined within the core of the system for a longer period of time. For the exponentially decaying and self-confined magnetic-field models, this confinement keeps the particles in regions of higher magnetic field strength, producing greater total synchrotron emission and therefore more stringent limits. Interestingly, for the uniform magnetic-field model, a lower diffusion coefficient instead leads to weaker limits. This counter-intuitive behaviour can be attributed to the geometric statistical averaging of the radial bins used in the likelihood evaluation. Annuli located further from the center enclose a vastly larger solid angle and thus contain a greater number of independent synthesized beams; this extensive averaging significantly reduces the empirical noise variance in the outer measurements. Conversely, the innermost bins of the profile suffer from greater variance due to containing very few independent samples. For a smaller diffusion constant, the $e^{\pm}$ particles do not diffuse outwards into these extended regions, forcing the concentrated central signal to be evaluated against the large statistical error of the innermost bins. Consequently, a higher diffusion constant transports the electrons into the outer annuli, which exhibit significantly lower variance, resulting in stronger constraints despite the more diffuse nature of the signal.

For a base comparison to previous limits, we have elected to include the conservative upper limits derived from Fermi-LAT $\gamma$-rays \cite{Donato_2012}. These results compare favorably to previous studies done using similar methodologies on MeerKAT Ultra-High Frequency (UHF) data \cite{Semane_2026} and the Australia Telescope Compact Array (ATCA) data \cite{Regis_2017}. While the results from this work are improved over the ATCA exclusion limits, they fall marginally short of the optimistic limits of the MeerKAT UHF results, and are instead comparable to the more conservative limits found by \citeauthor{Semane_2026} \cite{Semane_2026} with their inclusion of a flat-term correction to offset a slight negative bias in their residual background mean. The MeerKAT L-band spans a higher frequency range than that of the UHF band, and as such should have a slight advantage in terms of the RMS sensitivity of the observed data, due to the Galactic foreground sky temperature ($T_{\text{sky}}$) being lower at higher frequencies. However, through the inclusion of the free noise-scaling parameter $\alpha$ as well as the slightly larger independent beam sizes introduced through the covariance, our derived limits are slightly more conservative. 

While standard synchrotron radiation is typically brighter at lower frequencies ($S_\nu \propto \nu^{-\alpha_{\rm syn}}$, with spectral index $\alpha_{\rm syn}$), the synchrotron spectrum generated by DM annihilation is highly dependent on the WIMP mass. For high-mass WIMPs (i.e. $m_\chi \gtrsim 100$ GeV) annihilating through the $b\bar{b}$ channel, the injected $e^{\pm}$ pairs shift the peak of the synchrotron emission towards frequencies above $1$ GHz. Consequently, for these heavier candidates, observing at lower frequencies may actually yield weaker emission, highlighting the advantage of GHz-regime observations. However, either due to calibration uncertainties or systematics of the image, as well as a revised likelihood incorporating the uncertainties in the background estimation, the resulting annihilation limits of this work struggle to constrain the thermal models of higher mass WIMPs.

The decay bounds are presented in a similar manner to the annihilation case, again illustrating the impact of different magnetic-field models on the production channels over WIMP masses of $100\,{\rm GeV}$--$10$ TeV, as shown in the lower panel of Fig.~\ref{fig:excl}. For the decay analysis, we consider only $D_0 = 3.1\times 10^{28}\ \mathrm{cm^2\,s^{-1}}$. The essential difference between annihilation and decay lies in the dependence on the dark-matter density profile $\rho(r)$. For decay, the source term (Eq.~\ref{eq:decaysource}) scales linearly with $\rho(r)$ and the decay rate $\Gamma$. Owing to this linear $\rho$ dependence, the decay signal is substantially more diffuse and extended than in the annihilation case. Due to MeerKAT's maximum angular scale being $\approx 24'$ for our observational frequency range, any diffuse emission larger than this scale is effectively resolved out by the interferometer baseline coverage. The signal can still be observed, however it has to pass through an effective spatial filter created by the telescope itself. We imitate this effect in the model by subtracting the mean flux from the outermost bin in the largest image generated (in this case corresponding to $40'-45'$ away from the center) from the entire profile, which results in this bin representing the background after baseline subtraction, as done in the observed data by the interferometer. 

The spatial noise in wide-field PyBDSF-cleaned images is heavily correlated by the synthesized beam and exhibits significant non-Gaussian heavy tails due to deconvolution artifacts and primary beam corrections. While a purely uncorrelated Gaussian likelihood would underestimate the variance and lead to artificially strict limits, standard practice attempts to mitigate this by applying a penalty term that scales the likelihood by the number of pixels per synthesized beam. However, because the synthesized beam drops off as a Gaussian, adjacent pixels within a beam retain a fractional degree of statistical independence; a simple beam-scaling penalty often over-corrects for the correlation while failing to account for large-scale deconvolution artifacts. Therefore, as established in Sec.~\ref{sec:likelihood}, we construct the baseline covariance matrix $\mathbf{C}$ directly from the 2D spatial power spectrum of the residual map and absorb residual background uncertainties via the free noise-scaling parameter $\alpha$.

The methodological motivation for incorporating $\alpha$ purely into the covariance matrix $\alpha^2 \mathbf{C}$, rather than modifying the model subtraction term in the $\chi^2$, is the avoidance of severe parameter degeneracy. Specifically, the observed radial profile exhibits a localized negative flux dip in the innermost annular bins, with $I_{\rm obs} \approx -3\,\mu\text{Jy beam}^{-1}$. A potential alternative to account for this feature would be introducing an additive background or systematic bias profile $I_{\rm bkg}(\theta)$ directly into the predicted intensity, replacing $N_{\rm DM} I_{\rm syn}(\theta)$ with $N_{\rm DM} I_{\rm syn}(\theta) + N_{\rm bkg} I_{\rm bkg}(\theta)$. However, as the theoretical synchrotron profile $I_{\rm syn}(\theta)$ is sharply peaked at the center for the self-confined or exponential decay magnetic field profiles, $N_{\rm DM}$ and the background amplitude $N_{\rm bkg}$ become almost perfectly degenerate ($\rho \approx 0.99$ for the exponential profile). Such degeneracy destabilizes the likelihood and destroys the derived upper limits.

Handling systematic background uncertainties via the noise-scaling parameter $\alpha$ in $\alpha^2 C(\theta_a, \theta_b)$ completely eliminates this degeneracy, as the model comparison term remains strictly proportional to $N_{\rm DM}$. Nevertheless, this approach affects the sensitivity of the limits through scaling the entire covariance matrix by a single global factor $\alpha^2 \approx 18$ ($\alpha = 4.24$), which applies the same variance inflation to the outer annular bins, where synthesized beam correlations have already decayed and the noise is statistically well-behaved. As a result, the exclusion limits derived here are conservative relative to a framework that accounts for the scale-dependent decay of spatial correlations, since the statistical constraining power of the well-behaved outer bins is unnecessarily penalized. In future work, this can be refined without reintroducing parameter degeneracy by parameterizing the covariance matrix $C(\theta_a, \theta_b; \ell_{\rm corr})$ as an explicit function of the residual spatial correlation length $\ell_{\rm corr}$, derived from the 2D image auto-correlation function. This naturally inflates the variance specifically in the inner bins where the number of independent beam elements is small, while preserving the full statistical weight of the uncorrelated outer bins.

\section{Conclusions} \label{sec:conclusions}

We have presented the results obtained from observing the dSph galaxy Reticulum II using the L-band (856--1712 MHz) of MeerKAT. After an integration time of 6 hours, the data were reduced, calibrated, and imaged using \texttt{CARACal}, yielding a 6k pixel residual map with an RMS sensitivity of $4.8 \hspace{0.3em}\mu\text{Jy beam}^{-1}$. We utilized this image to constrain models of DM-sourced radio emission.

No significant excess of such emission was found above the background noise level. This implies that any potential DM signal must be comparable to or fainter than the RMS of the data, a result consistent with our null hypothesis. The image was constructed using a Briggs robust weighting of 0.5, providing a favorable balance between angular resolution and sensitivity.

We investigated the expected synchrotron radiation from $e^{\pm}$ populations produced via DM annihilation and decay, placing particular focus on the impact of different magnetic field models. The most optimistic scenario was represented by a uniform magnetic field, while a self-confined magnetic field provided a more conservative approach, and an exponentially decaying magnetic field yielded the most conservative limits, albeit physically less realistic. We further investigated the role of $e^{\pm}$ spatial diffusion, parametrized by the diffusion constant $D_{0}$. We demonstrated that variations in $D_{0}$ have divergent effects depending on the underlying magnetic field profile. This effect must be carefully considered in future analyses, as it dictates the stringency of the resulting limits.

To improve upon these constraints, several observational and analytical advancements can be considered. Deeper integrations can be explored, though the confusion floor of MeerKAT at 1.28 GHz was determined to be $\sim0.55 \, \mu\text{Jy beam}^{-1}$ \cite{Mauch_2020}, hence sensitivity improvements from raw integration time alone would be marginal. The increased baseline density and angular resolution of future arrays like the SKA could provide significant improvements in sensitivity. Additionally, targeting higher frequencies for a more sensitive RMS measure should prove beneficial, specifically the 1--3 GHz range for WIMP masses above 100 GeV \cite{regis2026probingfundamentalnatureparticle}. Finally, reducing astrophysical uncertainties, primarily through constraining the target's magnetic field profile (as the diffusive environment is strongly dependent on this property as well), will substantially improve the precision and robustness of the derived limits.

On the analytical front, refining the statistical treatment of spatial noise correlations offers a promising path forward. While our current implementation of the global noise-scaling parameter $\alpha$ successfully prevents the severe parameter degeneracy that would arise from additive central background templates, it results in overly conservative constraints in the outer annular bins. Generalizing the covariance matrix $C(\theta_a, \theta_b)$ to explicitly depend on the spatial auto-correlation length will allow future work to absorb localized central deconvolution artifacts, while restoring optimal sensitivity across the extended halo profile. Combined with refined DM halo properties and galaxy-specific diffusion models, these statistical developments will further improve methodologies for future radio upper limits studies.

\section*{Acknowledgements} \label{sec:acknowledgements}
    We would like to acknowledge the Centre for High Performance Computing (CHPC), South Africa, for providing the high-performance computing facilities and technical support necessary for the data reduction, imaging and analysis conducted in this work.
    This work was also supported by the South African Radio Astronomy Observatory (SARAO). The authors also acknowledge the use of data from the MeerKAT radio telescope, operated by SARAO, a facility of the National Research Foundation, an agency of the Department of Science, Technology and Innovation. Y.-Z. Ma acknowledges South Africa Research Chair Initiative (SARChI; Grant No. RCCA250325306198) from South Africa's Department of Science, Technology and Innovation and National Research Foundation (NRF), and other NRF Grants with No.~150580, No.~CHN22111069370, No.~ERC250324306141, and No. AUPP250310302114.

\bibliography{cleaned_bib}
\end{document}